\documentclass{emulateapj}
\usepackage{graphicx} 
  
\def\refjnl#1{{\rm#1}}
\def\apjl{\refjnl{ApJ}}                
\def\apjs{\refjnl{ApJS}}               
\def\ao{\refjnl{Appl.~Opt.}}           
\def\aap{\refjnl{A\&A}}                

\def\be{\begin{equation}} 
\def\ee{\end{equation}}

\def\gsim{\lower.5ex\hbox{\gtsima}} 
\def\lsim{\lower.5ex\hbox{\ltsima}} \def\gtsima{$\; \buildrel > \over \sim \;$} \def\ltsima{$\; \buildrel < \over \sim \;$} \def\prosima{$\; 
\buildrel \propto \over \sim \;$} \def\gsim{\lower.5ex\hbox{\gtsima}} 
\def\lsim{\lower.5ex\hbox{\ltsima}} 
\def\simgt{\lower.5ex\hbox{\gtsima}} 
\def\simlt{\lower.5ex\hbox{\ltsima}} 
\def\simpr{\lower.5ex\hbox{\prosima}}   
  
 \def\gtsima{$\; \buildrel > \over \sim \;$} 
\def\ltsima{$\; \buildrel < \over \sim \;$} 
\def\gsim{\lower.5ex\hbox{\gtsima}} 
\def\lsim{\lower.5ex\hbox{\ltsima}} 
\def\simgt{\lower.5ex\hbox{\gtsima}} 
\def\simlt{\lower.5ex\hbox{\ltsima}} 
\def\simpr{\lower.5ex\hbox{\prosima}}

\def\E3{{\cal E}_{\rm g}^{III}}

\def\Msun{\rm M_\odot}

\def\MSun{\rm M_\odot}

\def\M*{M_*}
\def\Z*{Z_*}
\def\L*{L_*}

\def\lgrb{$l{\rm GRBs}$}

\shorttitle{Habitability of the Universe through 13 billion years}
\shortauthors{Dayal et al.}

\begin{document}

\title{The habitability of the Universe through 13 billion years of cosmic time}
\author{Pratika Dayal\altaffilmark{1,2,3}, Martin Ward\altaffilmark{2} \& Charles Cockell\altaffilmark{4}}
\altaffiltext{1}{Kapteyn Astronomical Institute, University of Groningen, P.O. Box 800, 9700 AV Groningen, The Netherlands}
\altaffiltext{2}{Institute for Computational Cosmology, Department of Physics, Durham University, South Road, Durham DH1 3LE, UK}
\altaffiltext{3}{Institute for Advanced Study, Durham University, Palace Green Durham, DH1 3RL, UK}
\altaffiltext{4}{UK Centre for Astrobiology, School of Physics and Astronomy, University of Edinburgh, EH9 3HJ, UK}

\begin{abstract}
{The field of astrobiology has made tremendous progress in modelling galactic-scale habitable zones which offer a stable environment for life to form and evolve in complexity. Recently, this idea has been extended to cosmological scales by studies modelling the habitability of the local Universe in its entirety \citep{dayal2015, li2015}. However, all of these studies have solely focused on estimating the potentially detrimental effects of either Type II supernovae (SNII; explosions of stars more massive than 8 times the solar mass) or Gamma Ray Bursts (GRBs), ignoring the contributions from Type Ia supernovae (SNIa; explosions of stars between 3-16 times the solar mass) and active galactic nuclei (AGN; powered by accretion of gas onto a black hole). In this study we follow two different approaches, based on (i) the amplitude of deleterious radiation and (ii) the total planet-hosting volume irradiated by deleterious radiation. We simultaneously track the contributions from the key astrophysical sources (SNII, SNIa, AGN and GRBs) for the entire Universe, for both scenarios, to determine its habitability through 13.8 billion years of cosmic time. We find that SNII dominate the total radiation budget and the volume irradiated by deleterious radiation at any cosmic epoch closely followed by SNIa (that contribute half as much as SNII), with GRBs and AGN making up a negligible portion ($\lsim 1\%$). Secondly, as a result of the total mass in stars (or the total number of planets) slowly building-up with time and the total deleterious radiation density, and volume affected, falling-off after the first 3 billion years, we find that the Universe has steadily increased in habitability through cosmic time. We find that, depending on the exact model assumptions, the Universe is 2.5 to 20 times more habitable today compared to when life first appeared on the Earth 4 billion years ago. We find that this increase in habitability will persist until the final stars die out over the next hundreds of billions of years. } \\
\end{abstract}

\begin{keywords}
{Astrobiology, galaxies: stellar content, galaxies: quasars, methods:analytical, stars: gamma ray bursts, stars: supernovae}
\end{keywords}

\section{Introduction}
The field of cosmobiology \citep{bernal1952, dick1996} aims at understanding the habitability of the Cosmos in its entirety. This field naturally builds on the progress made in understanding the habitable zones around individual stars, the stellar habitable zones \citep[SHZ; ] []{huang1959, hart1979, kasting1993, rushby2013, kopparapu2013,guedel2014}, as well as galactic-scale habitable zones \citep[GHZ;][]{gonzalez2001, lineweaver2004, gowanlock2011, suthar2012, carigi2013, forgan2015}. The idea of habitability on a galactic scale is based on the premise that planets form around visible-light producing stars in regions that contain sufficient amount of elements heavier than Helium, by convention referred to as ``metals". However, complex multicellular surface-dwelling life on such planets can be adversely affected by deleterious radiation from a variety of astrophysical sources including: Type Ia Supernovae (SNIa; explosions of stars between 3-16 times the solar mass $\Msun$), Type II Supernovae (SNII; explosions of stars more massive than 8 $\Msun$), long-duration Gamma Ray Bursts (\lgrb; explosions of low-metallicity stars more than 20 $\Msun$) and Active Galactic Nuclei (AGN; powered by accretion of gas onto a central black hole). 
In terms of deleterious radiation, about 99\% of the total SN explosion energy \citep[$\simeq 10^{53}$ erg;][]{shklovskii1960, woosley2005, chevalier1981} is released as neutrinos. Although they have a very low interaction cross-section, SN exploding at a distance of 10 parsecs would produce $\simeq 10^4$ neutrino/nuclei recoil events per kilogram of tissue, potentially damaging it \citep{scharf2009}. Secondly most of the GRB energy \citep[$\simeq 10^{54}$ erg;][]{woosley2006}, is released as extremely high-energy gamma rays ($\gamma$-rays) while AGN produce extremely energetic X-rays  that can have a detrimental effect on habitability. Indeed, radiation from astrophysical events including \lgrb\, \citep{thorsett1995, dar1998, scalo2002, melott2004, thomas2005a} and SN \citep{ruderman1974, ellis-schramm1995,crutzen1996} have been postulated to be a cause of extinctions on the Earth. While there is currently no unequivocal evidence in the rock record linking these events to mass extinctions this may be a result of the difficulty of finding evidence in the rock record or the lack of events close enough to the Earth during the Phanerozoic, when large multi-cellular organisms have been present, to cause mass extinction. Nevertheless, models suggest that close astrophysical events near habitable worlds are likely to be deleterious to surface dwelling life by causing perturbations to atmospheric chemistry. While a number of {\it astrobiological} studies have attempted to estimate the effects of SNII and \lgrb\, in potentially rendering galactic-scale habitable zones inhospitable to complex life \citep[e.g.][]{clark1977, ellis-schramm1995, piran2014, piran2015, vukotic2016}, recent studies have substantially extended these models to the cosmobiological regime by calculating the limiting effects of SNII \citep{dayal2015} and \lgrb\, \citep{li2015} on the habitability of the entire local Universe using a sample of about 150,000 galaxies, with measured gas-phase metallicities, observed using the Sloan Digital Sky Survey (SDSS). 

However, any complete analysis of cosmological habitability must {\it simultaneously} quantify the potentially complex-life inhibiting consequences for all the key astrophysical sources (SNIa, SNII, \lgrb, AGN) in a holistic way, which is an unexplored approach that we tackle in this paper. It is important to note that - given their different progenitors - SN, \lgrb\, and AGN originate in galaxies of very different masses. Further, their relative numbers change as a function of redshift (or the age of the Universe) as successively larger galaxies build up with time according to the standard hierarchical model of galaxy formation. 

\begin{figure}
\center{\includegraphics[scale=0.49]{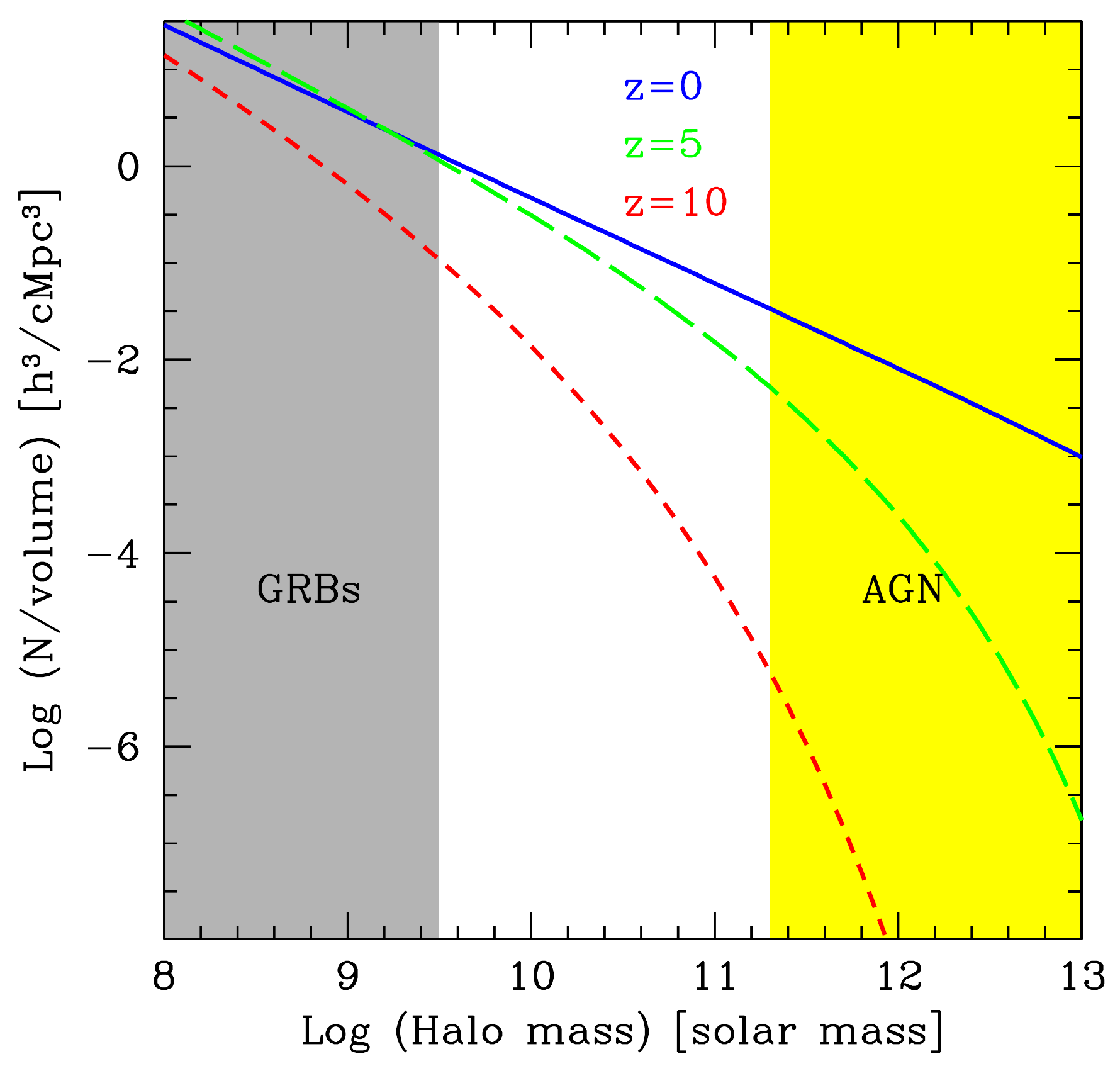}}
   \caption{The halo mass function showing the number of halos per unit comoving volume as a function of the halo mass. The short-dashed, long-dashed and solid lines show the results for $z=10,5$ and 0, corresponding to 0.5, 1.8 and 13.8 billion years after the Big Bang, respectively. As seen from this plot, the sources of potentially ``harmful radiation" continually increase through cosmic time as more and more halos build up. The shaded grey and yellow regions show the mass ranges affected by \lgrb\, and AGN feedback respectively, while any galaxy with star-formation will be affected by SN. As seen from this plot, the contribution of AGN rises most steeply through cosmic time, with the number density of low-mass $l{\rm GRB}$ hosting halos showing the least change. }
  \label{fig1}
\end{figure}

As shown in Fig. \ref{fig1}, \lgrb\, preferentially explode in the most numerous low-mass metal-poor galaxies with a halo mass $M_h \lsim 10^{9.5} \MSun$ and metallicity less than 30\% of the solar value \citep{savaglio2009, salvaterra2012, piran2014}; we neglect short-duration GRBs because of their negligible complex-life threatening effect \citep{piran2014}. On the other hand massive, and hence rarer, galaxies are expected to host AGN at any cosmic epoch. Combining the observed $M-\sigma$ relation, that links the black hole mass to the bulge mass \citep{haering2004}, with theoretical estimates of the ratio between the bulge mass and the dark matter halo mass \citep{vogelsberger2014} we derive a minimum AGN host halo mass of 
$M_h \gsim 10^{11.3} \MSun$ which is in agreement with expected values \citep[e.g.][]{ferrarese2000, mandelbaum2009, hickox2011}. However, intermediate mass galaxies with $M_h \simeq 10^{9.5}-10^{11.5} \MSun$ host the largest total amount of star formation, resulting in the largest number of SNII and SNIa. As seen from this figure, the number density of rare, massive galaxies hosting AGN increases by the largest amount between $z \simeq 10$ and 0, as the universe ages from being 500 million to 13.8 billion years old, with the number density of small ubiquitous galaxies hosting\, \lgrb\, increasing by the smallest amount. 

In this paper, we use observations of the AGN and star formation densities over the entire cosmic history to explore different models of habitability and ask, ``{\it which astrophysical source(s) dominate the potentially life-inhibiting radiation in the Universe, and how does this evolve through 13.8 billion years of cosmic time}"? Further, an increase in the total number of halos through time naturally implies an increase in the number density of each of these sources in an ageing Universe, resulting in increasing amounts of, and larger volumes affected by, deleterious radiation. Incorporating this understanding into our models, we ask: {\it when has the Universe been most habitable with respect to deleterious radiation threat to complex life?}

The cosmological parameters used in this work are: ($\Omega_{\rm m },\Omega_{\Lambda}, \Omega_{\rm b}, h, n_s, \sigma_8) = (0.2725,0.702,0.04, 0.7, 0.96, 0.83)$, consistent with the latest results from the {\it Planck} collaboration \citep{planck20132}.

\section{The astrophysical observations}
\label{sec_obs}
\begin{figure}
\center{\includegraphics[scale=0.498]{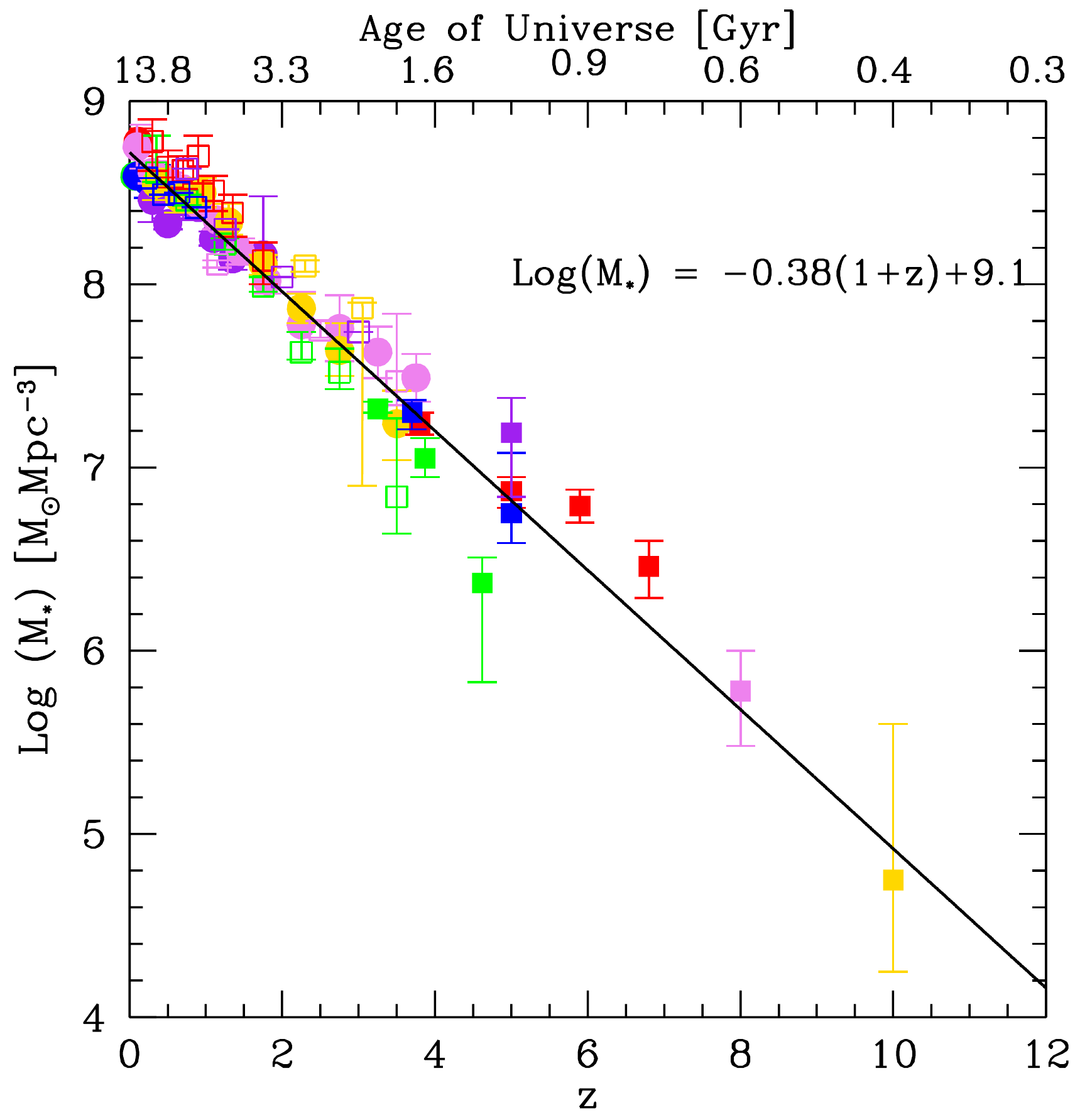}}
   \caption{The stellar mass per unit comoving volume (stellar mass density; SMD) as a function of $z$. Points show the stellar mass densities inferred from space and ground based observations by: \citet[][solid green circles]{li-white2009}, \citet[][solid red circles]{gallazzi2008}, \citet[][solid blue circles]{moustakas2013}, \citet[][solid purple circles]{bielby2012}, \citet[][solid violet circles]{perez-gonzalez2008}, \citet[][solid gold circles]{ilbert2013}, \citet[][empty green squares]{muzzin2013}, \citet[][empty red squares]{arnouts2007}, \citet[][empty blue squares]{pozzetti2010}, \citet[][empty purple squares]{kajisawa2009}, \citet[][empty violet squares]{marchesini2009}, \citet[][empty gold squares]{reddy2012}, \citet[][filled green squares]{caputi2011}, \citet[][filled red squares]{gonzalez2011}, \citet[][filled blue squares]{lee2012}, \citet[][filled purple squares]{yabe2009}, \citet[][filled violet squares]{labbe2013} and \citet[][filled gold square]{oesch2014}. We find that the $z$-evolution of the SMD is well fit by a power law: $Log(M_*) = -0.38(1+z)+9.1$.}
  \label{fig_smd}
\end{figure}

The habitability of the Universe at any cosmic epoch depends on the fraction of planets that can evolve unmolested by radiation potentially detrimental to the development of complex life such as energetic cosmic radiation, X-rays and $\gamma$-rays. To a first approximation, the total number of planets in a given volume scales with the total number of stars in that volume (the stellar mass density) and their metallicity. We use observations of the total stellar mass density (SMD, $M_*(z)$) shown in Fig. \ref{fig_smd} to infer that it grows smoothly with redshift as 
\begin{equation}
Log[M_*(z)/\Msun] = -0.38(1+z) + 9.1. 
\end{equation}
This relation shows that the SMD builds up very quickly at the earliest cosmic epochs as a progressively larger number of galaxies form and undergo star formation with the rate slowing at later times. We note that this observed SMD represents the total mass locked up in stars at any cosmic epoch, already including the mass lost due to explosions of massive stars\footnote{While the observed SMD includes the mass contribution from stellar remnants that do not host planets, they contribute $<1\%$ to the total SMD; we ignore this minor correction in our calculations.}. Given that Kepler data has shown planets can form around light-producing metal-rich stars with masses as low as $0.1 \Msun$\footnote{http://kepler.nasa.gov/Mission/discoveries/. See objects Kepler-42b,c and d.}, we use the Salpeter initial mass function \citep[IMF;][]{salpeter1955} - that describes the distribution of mass in a freshly formed stellar population- between $0.1$ and $100 \MSun$ throughout this paper. With this IMF, we calculate that most ($\sim 95\%$) of this mass is locked up in $\lsim 1\Msun$ stars with lifetimes ranging between about ten billion years (for a $1\Msun$ star) to about 50 billion years for a $0.1\Msun$ star \citep{chiappini1997}.

\begin{figure}
\center{\includegraphics[scale=0.498]{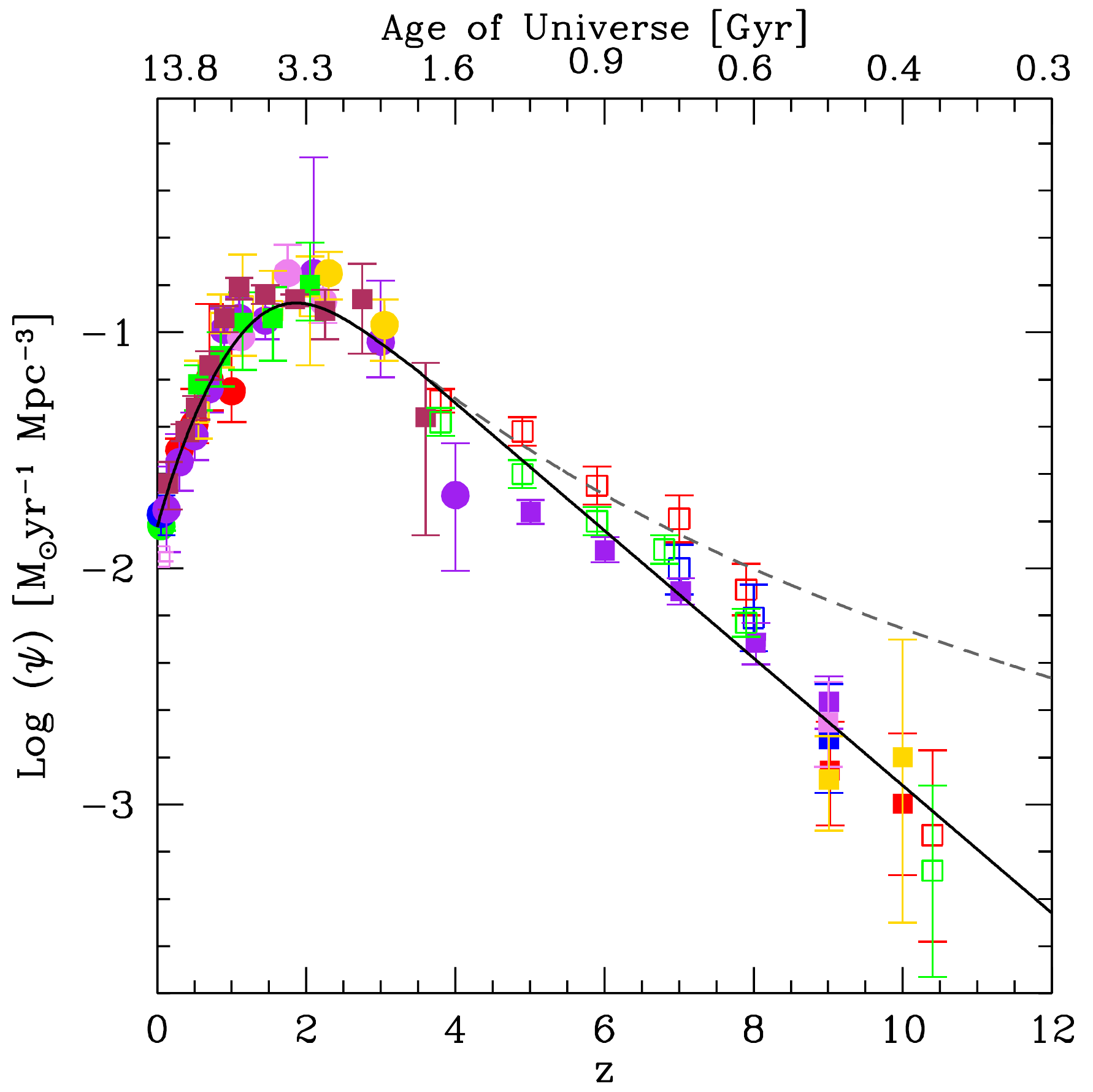}}
   \caption{The star formation rate per unit comoving volume (star formation rate density; SFRD) as a function of $z$ showing its build-up with time up to the peak at $z\simeq 2$, and the following decline. Points show the SFRD inferred from space and ground based observations by: \citet[][solid green circles]{wyder2005}, \citet[][solid red circles]{schiminovich2005}, \citet[][solid blue circles]{robotham2011}, \citet[][solid purple circles]{cucciati2012}, \citet[][solid violet circles]{dahlen2007}, \citet[][solid gold circles]{reddy2009}, \citet[][empty red squares]{bouwens2012a, bouwens2012b}, \citet[][empty blue squares]{schenker2013}, \citet[][empty purple squares]{sanders2003}, \citet[][empty violet squares]{takeuchi2003}, \citet[][empty gold squares]{magnelli2011}, \citet[][solid green squares]{magnelli2013}, \citet[][solid maroon squares]{gruppioni2013}, \citet[][solid red squares]{ellis2013}, \citet[][solid blue squares]{ishigaki2015}, \citet[][solid purple squares]{mcleod2015}, \citet[][solid violet squares]{mclure2013}, \citet[][solid gold squares]{oesch2014} and \citet[][empty green squares]{bouwens2015}. The dashed line shows the best-fit relation (based on low-$z$ data) inferred by \citep{madau2014}. However, the latest Hubble data shows a much sharper decline as seen from the $z \gsim 5$ data points. The solid line shows the best-fit  fiducial SFRD inferred in this work which is a combination of that given by \citet{madau2014} at $z \lsim 3.5$, and a power-law decline at higher $z$ (see Sec. \ref{sec_obs}). }
  \label{fig_sfrd}
\end{figure}

Next, we consider the redshift evolution of the ongoing star formation rate density (SFRD) that determines the SN and $l{\rm GRB}$ rate at any redshift. \citet{madau2014} have compiled the SFRD from a number of $z\simeq 0-9$ studies to show that this evolves as
\begin{equation}
\psi(z) = 0.015 \frac{(1+z)^{2.7}}{1+[(1+z)/2.9]^{5.6}} \,\,\,\,\,M_\odot \, {\rm yr^{-1} \, Mpc^{-3}}.
\end{equation}
As shown in Fig. \ref{fig_sfrd}, the SFRD increases by about two orders of magnitude over the 2 billion years between $z \sim 10.5$ and $z \sim 3$, reaches its peak at $z \sim 2$ and thereafter declines smoothly by about an order of magnitude till the present day. However, the latest datasets for high-$z$ galaxies complied using Hubble Space Telescope surveys \citep{ellis2013, mclure2013, oesch2014, ishigaki2015, bouwens2015, mcleod2015} show a much steeper evolution at the earliest cosmic epochs that can be fit by a simple power law at $z>3.5$ such that: 
\begin{equation}
\psi(z) = 1.12 \times 10^{-0.27(1+z)} \,\,\,\,\, M_\odot \, {\rm yr^{-1} \, Mpc^{-3}}
\end{equation}

We combine this information to yield the {\bf fiducial} SFRD (in $M_\odot \, {\rm yr^{-1} \, Mpc^{-3}} $) shown in Fig. \ref{fig_sfrd}, where:
\begin{eqnarray}
\psi(z\leq 3.5) & = & 0.015 \frac{(1+z)^{2.7}}{1+[(1+z)/2.9]^{5.6}} \nonumber \\
\psi(z>3.5) & = & 1.12 \times 10^{-0.27(1+z)} 
\label{fidsfrd}
\end{eqnarray}
In addition to this fiducial model, we also present results using the shallower SFRD proposed by \citet{madau2014} through-out the paper. 

Finally, the metallicity budget of the Universe slowly builds up with time with the (solar metallicity) Milky Way being a ``typical" $z \simeq 0$ galaxy. While metallicities of early galaxies are extremely hard to pin down, theoretical models tentatively indicate the stellar metallicities of $z\simeq 7$ galaxies to be of the order of $\sim 10-20\%$ of the solar value \citep[e.g.][]{dayal2013fmr}. We therefore make the reasonable simplifying assumption that the global stellar metallicity evolves as $Z_*(z) \simeq Z_* (z=0) [1+z]^{-1}$.

\section{The Model}
We start by calculating the number density of planets that depends both on the total stellar mass density, $M_*(z)$, available to host planets and the average metallicity at that epoch, $Z_*(z)$. \citet{buchhave2012} have found small terrestrial planets (radius $\lsim$ 4 times the earth radius) to be equally probable around stars with metallicities ranging between $0.25-2.5$ solar metallicity from Kepler data for 226 exoplanets and spectroscopy of their host stars. We therefore assume that the probability of finding terrestrial planets can be related to the stellar metallicity as $Z_*^\alpha$, where $\alpha$ is a free parameter. We explore its value in two limiting scenarios: $\alpha=0$ ($\alpha=1$) such that there is no (a strong) probability dependence of planets on the stellar metallicity. The total number of planets per unit volume ($N_p$) at any redshift can then be expressed as
\begin{equation}
N_p (z) \propto M_*(z)  Z_*^\alpha(z)
\end{equation}

We now use two different models, detailed below, to calculate the effects of deleterious radiation on the habitability of these planets. 

\subsection{Model 1: the influence of the total amount of deleterious radiation on habitability across cosmic time}
\label{app1}
This model is based on the assumption that the lower the intensity of deleterious radiation, the higher is the probability of habitability. We calculate the total deleterious radiation density of highly energetic cosmic radiation, X-rays and $\gamma$-rays in the Universe as a function of redshift, denoted by $\epsilon(z)$, including contributions from the four key astrophysical sources: SNII, SnIa, GRBs and AGN. The cosmic-averaged probability of habitability at any redshift, $P_{hab1}(z)$, can then be expressed as

\begin{equation}
P_{hab1}(z)  \propto \frac{M_{*}(z) Z_*^\alpha(z)} {\epsilon(z)},
\label{phab}
\end{equation}
where the numerator describes the total number density of planets while the denominator determines the fraction that would be habitable. Further, $\alpha$ can have a value of either 0 or 1 as noted above. We normalise $P_{hab1}(z)$ to its value at $z=0$ in order to remove the proportionality constant so that
\begin{equation}
\frac{P_{hab1}(z)}{P_{hab1}(0)}  = \frac{M_*(z)}{M_*(0)} \bigg(\frac{Z_*(z)}{Z_*(0)}\bigg)^\alpha \frac{\epsilon(0)}{\epsilon(z)}.
\label{phab1_rat}
\end{equation}

We now describe the calculation for $\epsilon(z)$. For the Salpeter IMF used in this work, the SNII rate density is related to $\psi(z)$ as:
\begin{equation}
\label{sn2}
rSNII(z) = \frac{\psi(z)}{134 \MSun} [{\rm yr^{-1} \, Mpc^{-3}}].
\end{equation}

Further, roughly 10\% of all stars between $3-16 M_\odot$ explode as SNIa \citep[e.g.][]{matteucci1986} yielding a SNIa rate density of
\begin{equation}
\label{sn1}
rSNI(z) = \frac{\psi(z)}{380 \MSun} [{\rm yr^{-1} \, Mpc^{-3}}]
\end{equation}

\begin{figure}
\center{\includegraphics[scale=0.49]{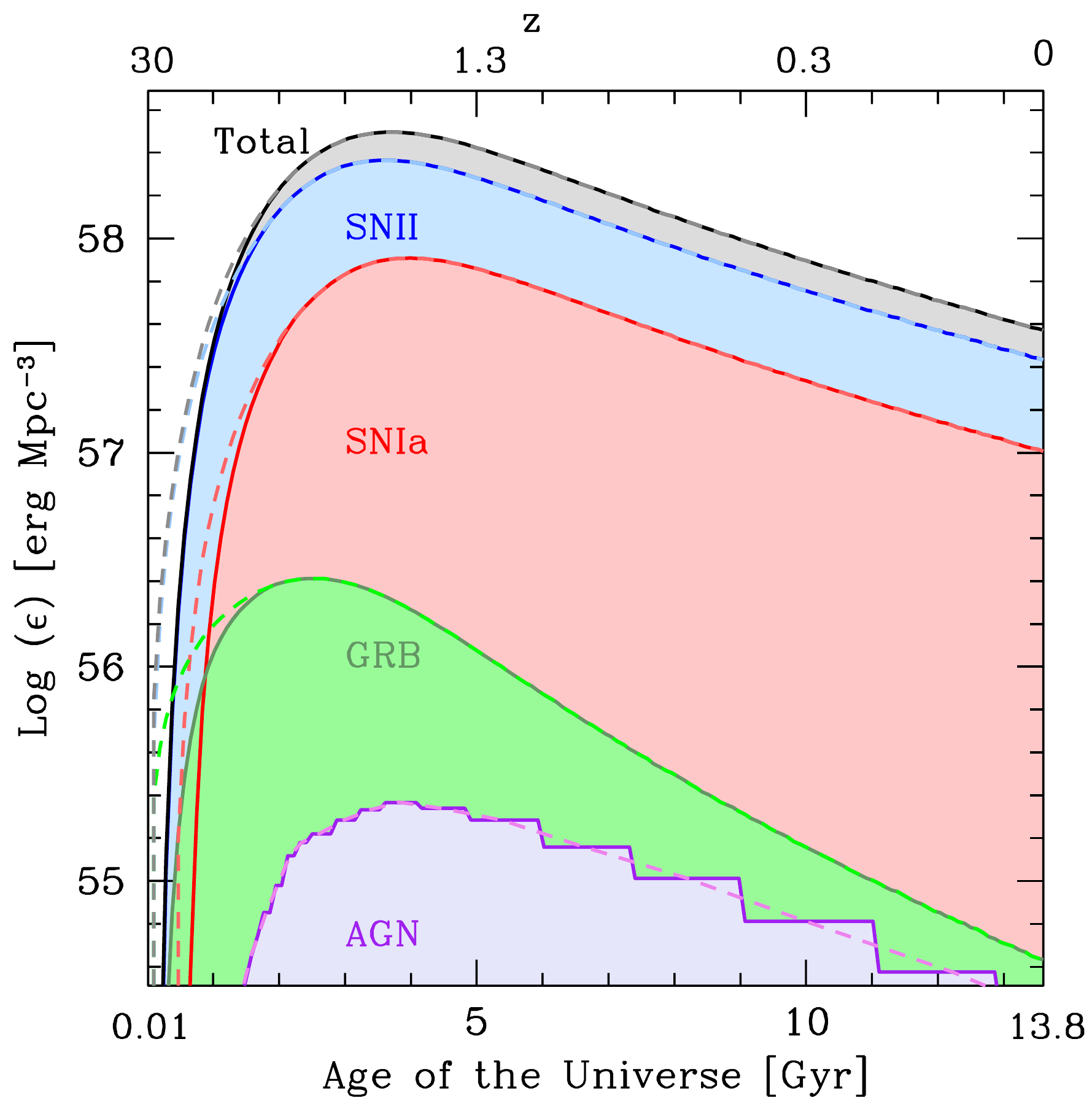}}
   \caption{The total deleterious radiation density as a function of the age of the Universe. Shaded violet, green, red and blue regions show the contributions from AGN (X-rays), \lgrb ($\gamma$-rays), SNIa and SNII (high energy neutrinos) respectively using the {\it fiducial SFRD model}; dashed lines show results using the SRFD measured by \citep{madau2014}. The solid black and dashed grey lines show the total radiation density, summing over all sources from the fiducial SFRD and \citet{madau2014} models, respectively. As seen, the total energy density is dominated by SNII at all cosmic epochs, closely followed by SNIa. The contribution from \lgrb\, (AGN) is lower by 2 (3) orders of magnitude at any $z$. Further, while star formation is well underway at $z \simeq 20$, the first SNIa appear as late as $z \simeq 10.5$ due to the delay of 370 Myrs between star formation and the first SNIa explosions.  }
  \label{fig_radden}
\end{figure}

Similarly, the $l{\rm GRB}$ rate can be related to SFRD as (Salvaterra et al. 2012)
\begin{equation}
\label{grb}
rlgrb(z) = \frac{0.24} {10^{-9}} f_b (1+z)^{1.7} \psi(z) [{\rm yr^{-1} \, Mpc^{-3}}],
\end{equation}
where the factor $f_b = 181$ accounts for the fact that GRBs are beamed with a typical opening angle of about 6 degrees, which results in about 180 ``hidden" GRBs for each one that is detected \citep{frail2001, ghirlanda2013}.

Whilst the SNII and GRB rate density simply trace the SFR, SNIa explosions are delayed by about 370 Myrs (which is when stars of a minimum mass of $3\MSun$ explode). These SN and GRB rate densities are converted into the {\it total number density} by multiplying by $\Delta t$, which is the time interval between two successive $z$ steps. Finally, we use the AGN X-ray luminosity density for sources with an X-ray luminosity greater than $10^{42} {\rm erg \, s^{-1}}$ for $z =0.01$ to $z=4.9$ \citep{ueda2014} and assume the value to remain unchanged from $z = 4.9$ to higher redshifts to obtain an upper limit on the total AGN deleterious radiation density. 

Then, the total deleterious radiative energy density at any $z$ can be expressed as
\begin{eqnarray}
\label{emm}
\epsilon(z) =  (rSNII(z) \times \Delta t \times 10^{53}) \nonumber\\
+ (rSNI(z) \times \Delta t \times 10^{53}) \nonumber\\
 + (rlgrb(z) \times \Delta t \times 10^{54}) \nonumber \\
 + (\epsilon_{agn} \times t1) 
\end{eqnarray}
where we assume AGN to radiate at the peak energy for as long as $t1=100$ Myr. 

The cumulative value of $\epsilon(z)$ and the contributions from the 4 key astrophysical sources are shown in Fig. \ref{fig_radden}. By construction, the SNII and $l{\rm GRB}$ energy density track the SFRD, showing an increase for the first 3 billion years, thereafter showing a decline; the radiation energy density decreases with redshift as $(1+z)^4$, both due to the expansion of the Universe, as well as cosmological redshift which results in the radiation from ongoing SN/GRBs/AGN dominating at any epoch. Whilst following the same qualitative trend, SNIa are delayed by 0.37 billion years, a timescale too small to be discernible on the figure. 

First, we find that {\it SNII dominate the total radiation density of the Universe at any epoch} with the SNIa contribution consistently being a factor of 2 lower, as expected from Eqns. \ref{sn2} and \ref{sn1}. Secondly, despite the upper limit of assuming GRBs to form in every galaxy with a mass lower than $10^9\MSun$ at any cosmic epoch, their contribution to the total radiation density is about a hundred times smaller than that from SNII. Thirdly, despite assuming a constant luminosity density for AGN at $z>4.9$, given the fact that only the largest and rarest galaxies can host AGN, their total radiation density is about a thousand times smaller than that from SNII at any epoch. Finally, we find that the two SFRD laws (fiducial versus Madau \& Dickinson 2014) used do not have any sensible impact on the total radiation density - this is only to be expected given they are only different in the first few hundred million years of the Universe.

\subsection{Model 2: linking habitability to the volume affected by deleterious
radiation across cosmic time}
\label{app2}
In this model we explore a scenario where the probability of habitability is inversely proportional to the fraction of total volume occupied by planets affected by deleterious radiation; this approach has previously been used to study the habitability of about 130,000 galaxies at $z=0$ \citep{dayal2015}. The probability of habitability in this case is expressed as

\begin{equation}
P_{hab2}(z) \propto \frac{M_{*}(z) Z_*^\alpha(z)} {f_{irr}(z)},
\label{phab2}
\end{equation}
where we again assume the total number of planets to scale with the SMD and stellar metallicity at any redshift as expressed by the numerator; as in Model 1, we explore values of $\alpha = 0$ and 1, representing no dependence and a strong metallicity dependence, respectively. Further, $f_{irr}(z) = V_{rad}(z)/V_{tot}(z)$ represents the fraction of the total volume hosting planets affected by deleterious radiation. Here, $V_{tot}$ is the total volume that hosts stars at a given redshift and $V_{rad}$ accounts for the total volume affected by deleterious radiation from all 4 astrophysical sources at that $z$. 

Making the reasonable assumption that the density of any galaxy is roughly 200 times the critical density of the Universe at that redshift ($\rho_c(z)$), the total volume occupied by all stars can be written as $V_{tot} = M_*(z) V_{obs} [200 \rho_c(z)]^{-1}$. Here $\rho_c(z) = 3 H(z)^2 [8\pi G]^{-1}$ where $H(z)$ is the Hubble parameter at the redshift considered and $G$ is the universal gravitational constant. The factor $V_{obs}$ accounts for the total volume that is observed to infer the stellar mass density - this is required to turn the stellar mass density into a stellar mass at any given $z$.

The total volume affected by deleterious radiation is the volume affected per astrophysical (SNIa/SNII/GRB/AGN) event times the number of events at that $z$ so that 
\begin{eqnarray}
\label{vrad}
V_{rad} (z) & = & Q V_{obs} [ rSNII(z) \times \Delta t + rSNI (z) \times  \Delta t  \nonumber\\
& & + rlgrb(z) \times  \Delta t + N_{agn}] \nonumber\\
V_{rad} (z) & = & Q V_{obs} r_{tot}
\end{eqnarray}

\begin{figure}
\center{\includegraphics[scale=0.49]{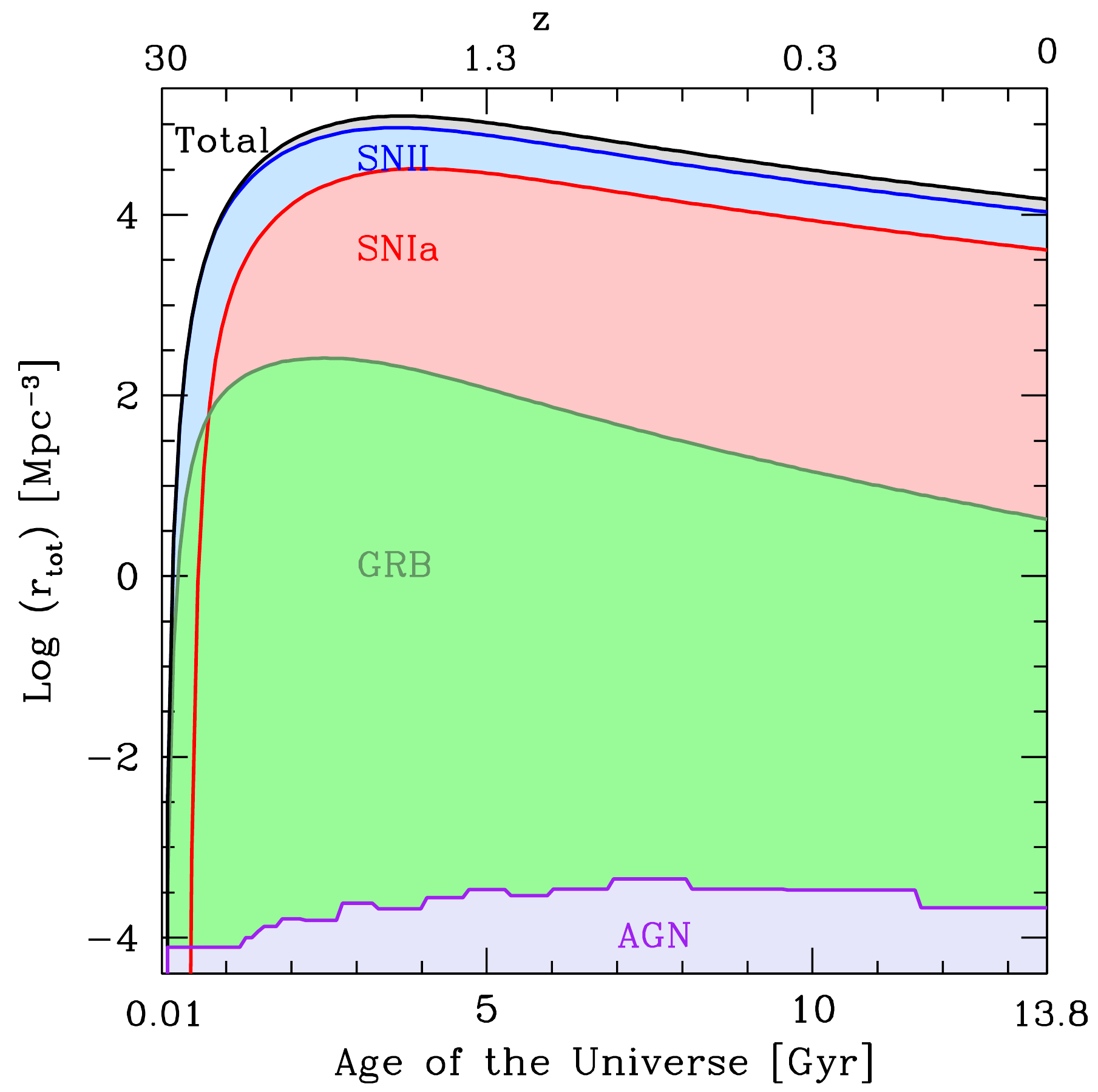}}
   \caption{We show $r_{tot}$, a proxy for the total volume hosting stars that is affected by deleterious radiation, as a function of the age of the Universe. Shaded violet, green, red and blue regions show the volumes affected by AGN (X-rays), \lgrb ($\gamma$-rays), SNIa and SNII (high energy neutrinos) respectively using the {\it fiducial SFRD model}. The solid black line represents the total volume affected, summing over all sources. As seen, $r_{tot}$ is dominated by SNII at all cosmic epochs, closely followed by SNIa. The contribution from \lgrb (AGN) is lower by 3 (8) orders of magnitude at any $z$. }
  \label{fig_vol}
\end{figure}

where we have made the simplifying assumption that, irrespective of its nature, each astrophysical event irradiates the same volume ($Q$) with deleterious radiation and $r_{tot}$ represents the number density of events producing deleterious radiation at a given $z$; we multiply the rate of astrophysical events by $V_{obs}$ to convert the SN, GRB and AGN rate densities \citep{ueda2014} into the total number of events. Further, $\Delta t$ is the time interval between two successive $z$ steps and the values of all other parameters are as in Eqns. \ref{sn2} to \ref{emm}. Naturally, $V_{obs}$ factors out in $f_{irr}$ so that $P_{hab2}(z)$ can be written as
\begin{eqnarray}
P_{hab2}(z) \propto \frac{M_{*}^2(z) Z_*^\alpha(z)} {Q  r_{tot} 200 \rho_c(z)}, 
\label{phab3}
\end{eqnarray}
which is again normalised to the value at $z=0$ to remove the constant of proportionality as
\begin{eqnarray}
\frac{P_{hab2}(z)}{P_{hab2}(0)}  = \frac{M_*^2(z)}{M_*^2(0)} \bigg(\frac {Z_*(z)}{Z_*(0)}\bigg)^\alpha \frac{[r_{tot}(0)]}{[r_{tot}(z)]} \frac{\rho_c(0)}{\rho_c(z)} 
\label{phab3_rat}
\end{eqnarray}

From Fig. \ref{fig_vol} we see that $r_{tot}$, which is a proxy for the total stellar volume irradiated by deleterious radiation (Eqn. \ref{vrad}), is {\it dominated by SNII with the volume affected by SNIa being a factor of 2 lower at all $z$}. Despite assuming (the upper limit of) every galaxy with a mass lower than $10^9\MSun$ hosting GRBs at any $z$, their contribution to the total volume irradiated is about three orders of magnitude lower than SN. Finally, despite assuming a constant number density for AGN at $z>4.9$, their contribution to the total stellar volume irradiated is about ten million times smaller than that from SN. Given that the two SRD laws used in this work (fiducial versus Madau \& Dickinson 2014) are only different in the first few hundred million years, we only show results for the fiducial model here. 

\begin{figure}
\center{\includegraphics[scale=0.49]{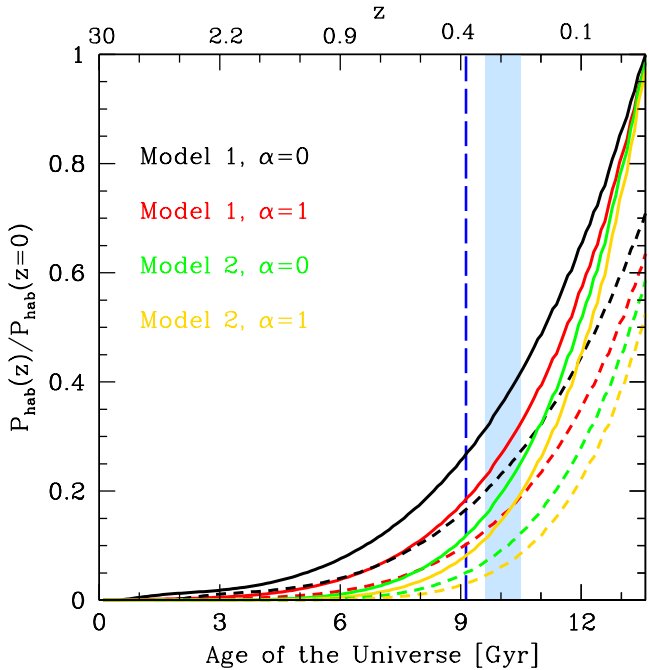}}
   \caption{The probability of habitability of the Universe as a function of its age, normalised to the present day. The solid black and red lines show the probability using the deleterious radiation energy density argument (Model 1) for no ($\alpha =0$) and maximum ($\alpha=1$) metallicity dependence, respectively. The solid green and gold lines show the probability using the volume irradiated by deleterious radiation (Model 2) for no ($\alpha =0$) and maximum ($\alpha=1$) metallicity dependence, respectively. The vertical dashed line at 9.1 billion years shows the formation of the Earth, with the dashed blue region showing the range when life is first though to have appeared on the Earth, 3.2 to 4.1 billion years ago. While differing in amplitude, the 4 different models explored all point towards habitability increasing with the age of the Universe. We find that the Universe was 5-30\% as habitable as today when the Earth formed, and between 5-40\% as habitable when life first appeared. Finally, dashed curves show the probability of habitability for each of the models factoring in a delay of 1.4 billion years, using the Earth as a template, for the appearance of complex life.}
  \label{fig_phab}
\end{figure}

\section{Results and discussion}
Now that we have calculated the total deleterious radiation density (model 1; Sec. \ref{app1}) and the total stellar volume irradiated by deleterious radiation (model 2; Sec. \ref{app2}), we use Eqns. \ref{phab1_rat} and \ref{phab2} to calculate the habitability of the Universe as a function of the its age, normalised to the present day - the results of these calculations are shown in Fig. \ref{fig_phab}. 

As shown, the habitability for both the models considered - either based on the amplitude of the deleterious radiation density or the total stellar volume irradiated - rises through time. The habitability rises the fastest for model 1, based on the deleterious radiation density, neglecting metallicity dependence (i.e. $\alpha=0$). Including metallicity dependence introduces a factor of $[1+z]^{-1}$ into Eqn. \ref{phab} decreasing the probability of habitability. In model 2, based on the total stellar volume irradiated by deleterious radiation, the probability of habitability depends on $M_*^2$ (see Eqn. \ref{phab3}) leading to a slower rise in its value. Again, the probability is higher in this case neglecting metallicity dependence. The rise of habitability is a combination of two effects: (i) the stellar mass density (i.e. density of planets) continually builds up through cosmic time; (ii) the deleterious energy radiation density and volume irradiated decrease as star formation winds down after the first 3 billion years of the Universe. 

Given the indiscernible difference in the radiation density between the two SFRD relations in the first few hundred million years, we only show results for the fiducial SRFD law. The two models presented bracket the probability of habitability and range between the Universe being $5-30\%$ as habitable as today when the Earth formed 4.46 billion years ago \citep{dalrymple2001}. We also find that the Universe was between $10-40\%$ as habitable as today when life is understood to have first appeared on the Earth 3.2 to 4.1 billion years ago \citep{bell2015,rasmussen2000}. Finally, we calculate the probability of habitability accounting for the delay (of about $1.4$ billion years) between the formation of planets and the appearance of life using the Earth as a template \citep[see also][]{lineweaver2004}. Factoring in this delay shifts all the habitability curves to later times resulting in a Universe that was $5-30\%$ as today as when life first appeared on the Earth. These results imply that, considering the different models explored, the Universe today is about 2.5 to 20 times more habitable today as compared to a time 4 billion years ago when life first appeared on the Earth. 

To summarise, we answer two of the fundamental questions regarding the trends and sources that determine habitability across 13.8 billion years of cosmic time: firstly, we find that independent of redshift, SNII are the key astrophysical source that dominate the potentially life-inhibiting radiation in the Universe, closely followed by SNIa. Secondly, we find that the habitability of the Universe as a whole has been steadily increasing across cosmic time. Neglecting the delay between the formation of planets and appearance of life of about 1.4 billion years, we find that the Universe has reached the maximum probability of habitability at the present day; however, accounting for this delay the habitability of the Universe is still building up towards its maximum value. 
Extending this argument to the future, we envisage a Universe that will continually increase in habitability, till the last stars run out of fuel sometime over the next hundreds of billion years. 

We end by noting the main caveats involved in this work. In this first attempt to understand habitability simultaneously accounting for the main astrophysical sources of deleterious radiation, we have assumed both the radiation and stellar mass densities to be homogeneously distributed throughout the Universe, presenting an `integrated' view on the habitability with respect to the minimum radiation threat. However, given the different environments they reside in, $l{\rm GRBs}$ and AGN would be expected to only affect the habitability of the least and most massive halos at any epoch, respectively. We have neglected the fact that the star formation rate generally scales with the halo mass at high-redshifts \citep{dayal2009} implying lower-mass galaxies would be more hospitable. On the other hand, the star formation rate trend reverses at lower-redshifts where the most massive ellipticals show quenched star formation - indeed, in a previous work \citep{dayal2015}, we have shown that massive ellipticals can host ten-thousand times as many hospitable planets as the Milky Way, making them the most probable cradles of life in the local Universe. While recent state-of-the-art simulations of the Milky Way are now able to resolve star particles of $10 \Msun$ \citep{bedorf2014}, such calculations are currently only possible for individual galaxies. In future works, we will aim to use large-scale cosmological simulations to model the ``galactic zone of influence" of each of these sources in galaxies of different masses through cosmic time; modelling stellar habitable zones using large-scale cosmological simulations remains implausible, at least for the foreseeable future. Finally, we have used an `average' fit to the SMD (Fig. \ref{fig_smd}) ignoring the observational dispersion (of about a factor of 5) most pronounced for $z<1$.

\section*{Acknowledgments} 
PD acknowledges support from the European Commission's CO-FUND Rosalind Franklin program and thanks the Adisson Wheeler fellowship and the Institute for Advanced Study at Durham University. PD thanks ``PACIFIC 2015" for a wonderful environment and the Gump Station for their hospitality, where this work was carried out. PD thanks Carlton Baugh, Carlos Frenk, Anupam Mazumdar and Ruben Salvaterra for insightful discussions. Finally, PD deeply thanks Emma McIntyre and Matilda Strang for inviting her to the Shona Project's Pilot weekend where this work was finished, and Fay Stevens, Florence Devereux, Paul Kinderseley, Ceylan Hay, Johanna McDorwuff, Frank McElhinney, Jo Ray and Erin Busswood for putting this work into perspective.


\bibliographystyle{apj}
\bibliography{ms}

\label{lastpage} 
\end{document}